\documentclass{article}

\usepackage[preprint,nonatbib]{neurips_2021}

\usepackage[utf8]{inputenc} 
\usepackage[T1]{fontenc}    
\usepackage{hyperref}       
\usepackage{url}            
\usepackage{booktabs}       
\usepackage{amsfonts}       
\usepackage{nicefrac}       
\usepackage{microtype}      
\usepackage{xcolor}         
\usepackage{graphicx}
\usepackage{todonotes}

\title{Lhotse: a speech data representation library for the modern deep learning ecosystem}

\author{%
  Piotr Żelasko\thanks{Center for Language and Speech Processing and Human Language Technology Center of Excellence.} \\
  Johns Hopkins University \\
  Baltimore, MD 21216 \\
  \texttt{piotr.andrzej.zelasko@gmail.com} \\
   \And
   Daniel Povey \\
   Xiaomi \\
   China \\
   \texttt{daniel.povey@gmail.com} \\
   \AND
   Jan "Yenda" Trmal$^*$ \\
   Johns Hopkins University \\
   Baltimore, MD 21216 \\
   \texttt{jtrmal@gmail.com} \\
  \And
   Sanjeev Khudanpur$^*$ \\
   Johns Hopkins University \\
   Baltimore, MD 21216 \\
   \texttt{khudanpur@jhu.edu} \\
}

\begin{document}

\maketitle

\begin{abstract}
  Speech data is notoriously difficult to work with due to a variety of codecs, lengths of recordings, and meta-data formats. We present Lhotse, a speech data representation library that draws upon lessons learned from Kaldi speech recognition toolkit and brings its concepts into the modern deep learning ecosystem. Lhotse provides a common JSON description format with corresponding Python classes and data preparation recipes for over 30 popular speech corpora. Various datasets can be easily combined together and re-purposed for different tasks. The library handles multi-channel recordings, long recordings, local and cloud storage, lazy and on-the-fly operations amongst other features. We introduce Cut and CutSet concepts, which simplify common data wrangling tasks for audio and help incorporate acoustic context of speech utterances. Finally, we show how Lhotse leverages PyTorch data API abstractions and adopts them to handle speech data for deep learning.
\end{abstract}

\section{Introduction}
\label{sec:intro}

The last 30 years have brought an abundance of speech recordings usable in artificial intelligence (AI) research. English alone has more than 100.000 hours of recorded speech freely available with the recent releases of LibriSpeech~\cite{panayotov2015librispeech}, Multilingual LibriSpeech~\cite{pratap2020mls}, LibriLight~\cite{kahn2020libri}, GigaSpeech~\cite{chen2021gigaspeech}, SPGISpeech~\cite{o2021spgispeech}, The People's Speech~\cite{galvez2021people}, and others.
However, speech data is notoriously difficult to work with for machine learning practitioners. Recordings of speech come in many flavors: as isolated utterances in separate files (e.g., LibriSpeech~\cite{panayotov2015librispeech}); long, continuous recordings of podcasts and conversations (e.g., GigaSpeech~\cite{chen2021gigaspeech}); or even multi-channel recordings from multiple microphone arrays (e.g., AMI~\cite{mccowan2005ami}, CHiME-6~\cite{watanabe2020chime}). Audio is encoded with a variety of codecs, both common (e.g., PCM, FLAC, OPUS) and obscure (e.g., sphere, shorten). The meta-data, used as supervision for model training, comes with a different schema for each dataset, usually as raw text files, TextGrid, XML, JSON, or others.

Many of these issues were addressed by the Kaldi toolkit~\cite{povey2011kaldi}. Kaldi introduced a concept of data directories, a standard representation for all datasets where multiple files describe different aspects of data, such as the audio file location, the speech segments, the transcript, the speaker, the utterance duration, and so on. This format has been adopted by other speech libraries such as ESPnet~\cite{watanabe2018espnet} and Espresso~\cite{wang2019espresso}. However, Kaldi was designed with speech experts in mind, and its learning curve is considered steep by speech students, non-speech folk, and even some within the speech community. Indeed, reaching a broader audience interested in speech modelling is one of our core motivations for the work at hand. 

In this paper we present Lhotse, a speech data representation library that takes the lessons learned from Kaldi and brings them into the modern deep learning ecosystem. 
Lhotse is one of the three libraries that constitute the next-generation Kaldi framework. The remaining two are \textit{k2} for differentiable, GPU-accelerated weighted finite state automata (WFSA) algorithms~\cite{k22021}; and \textit{Icefall} that contains simple, reproducible recipes for training and evaluating speech models~\cite{icefall2021}. These projects are licensed with the Apache 2.0 licence. 

\section{Core concepts in Lhotse}
\label{sec:main_concepts}

\begin{figure}
    \centering
    \includegraphics[width=\linewidth]{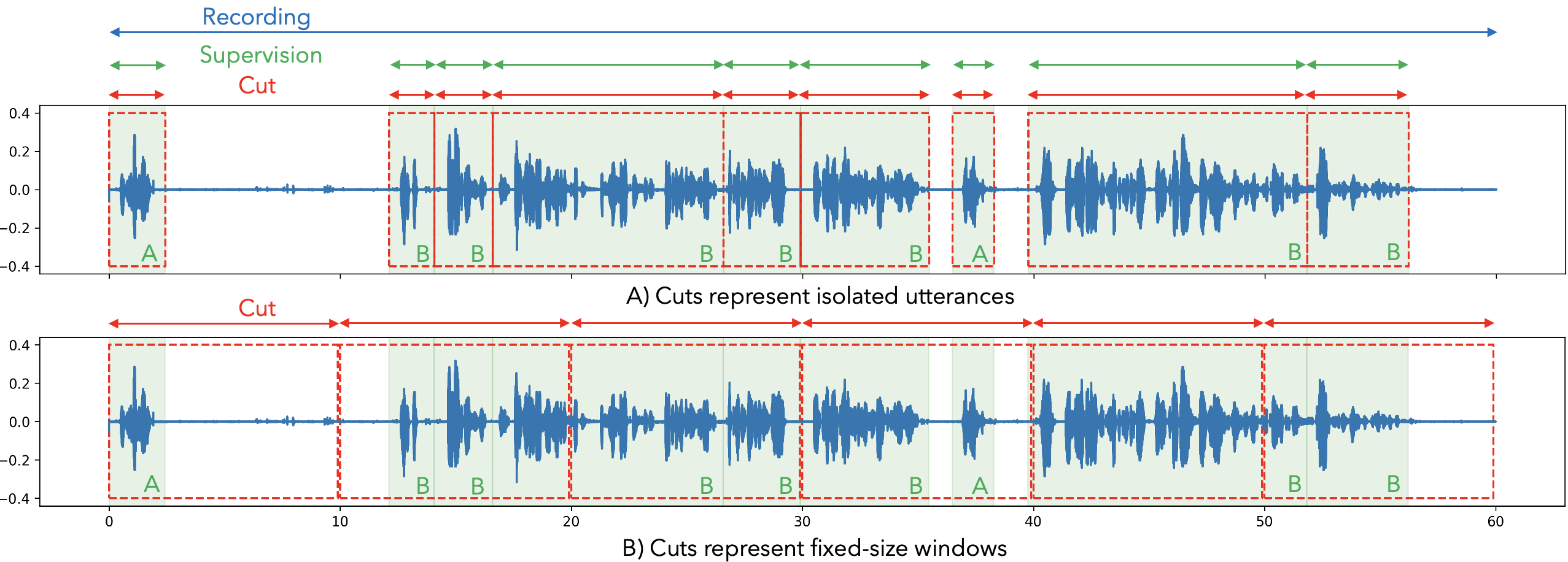}
    \caption{Recording, SupervisionSegments, and Cuts illustrated on waveform plots of a conversation with speakers A and B. The upper plot (A) shows cuts matching supervision boundaries in an "isolated utterances" scenario, e.g., for speech recognition. The bottom plot (B) shows how cuts are used to traverse the same recording in windows, useful in voice activity detection or speech diarization. Note that with different window sizes and offsets there are many ways to construct the training data.}
    \label{fig:lhotse_manifests}
\end{figure}

Lhotse defines speech manifest schemas for representing recordings and all of their associated meta-data. These manifests are mapped into Python objects, with their own set of methods for easy manipulation of speech data. Like Kaldi, Lhotse provides standard data preparation recipes for publicly available corpora (over 30 at the time of writing), and utilities to download them, if possible. Lhotse supports importing Kaldi data directories and exporting manifests to Kaldi.

\textbf{Basic data description: recording, supervision, and feature manifests.} There are four types of manifests within Lhotse: \texttt{Recording}, \texttt{SupervisionSegment}, \texttt{Features}, and \texttt{Cut}. \texttt{Recording} abstracts away from the physical location of the audio data and allows to read short and long, single- and multi-channel audio from one or more files, stored on a local disk, remote URL or a cloud storage service, possibly encoded in different formats, with a single API: \texttt{Recording.load\_audio()}. \texttt{SupervisionSegment} denotes a time span in the \texttt{Recording} that has some associated meta-data usable for supervised training: typically, it would be a speech segment with properties such as start, duration, speaker ID, gender, age, transcript, and possibly more depending on the corpus. 
\texttt{Features} is an optional manifest that helps to work with pre-computed features for model training or inference. It contains basic information such as tensor shape, frame shift, corresponding \texttt{Recording} ID, and so on. Similarly to \texttt{Recording}, it abstracts away from the storage mechanism and allows to read/write to a local file-system, a cloud storage, and more. Lhotse supports both on-the-fly and pre-computed feature extraction for greater versatility and leverages an efficient feature compression engine \texttt{lilcom}~\cite{lilcom2021}.

\textbf{Flexible training example construction with \texttt{Cut} manifests.} The core contribution of Lhotse is the concept of \texttt{Cut}. \texttt{Cut} may be viewed as a "window" with a certain offset and duration in a \texttt{Recording} that has zero or more \texttt{SupervisionSegment}s. As such, \texttt{Cut} ties together audio and meta-data with exact timing, without any assumptions about the number of utterances or speakers. \texttt{Cut} provides greater data preparation flexibility than was possible with Kaldi and other speech toolkits--it allows to construct training examples with additional acoustic context for each utterance (e.g., background noises in a telephone conversation that could help the network adapt) or even additional speech context (e.g., modelling contextual dependencies between utterances in a podcast). Notably, most operations performed on cuts are lazy -- whether it's mixing, truncation, padding, or augmentation, Lhotse reads only as much audio data as needed to deliver the end result.

\textbf{Universal speech corpus representation.} To further simplify data manipulation, Lhotse introduces an object representing a collection of cuts called \texttt{CutSet}. Ultimately, all speech corpora are represented as \texttt{CutSet}s for training or batch inference. \texttt{CutSet} has over 30 methods that simplify executing common audio \textit{data wrangling} tasks, such as padding, sub-setting, truncating, extracting features, or visualizing and listening to in Jupyter notebooks. \texttt{CutSet} also supports data augmentation that correctly adjusts the relevant meta-data, such as speech segments duration: mixing with other sound sources (babble, noise, music), perturbing speed, tempo, or volume, resampling, and so on. Finally, \texttt{CutSet} has a few methods for converting long recordings into shorter cuts suitable for different training schemes: isolated utterances (typical speech recognition setup), constant-sized windows (typical voice activity detection or speech diarization setup), or extracting just the background noises (useful for mixing with out-of-domain speech as training data augmentation). The isolated utterance and constant-sized windows schemes are visualized in Figure~\ref{fig:lhotse_manifests}. Finally, the \texttt{CutSet} concept makes it very easy to re-purpose the same data for different tasks, and even multi-task training schemes, as all the meta-data is stored next to each other. Combining different datasets together is also straightforward.

\textbf{What Lhotse is not.} To avoid confusion, it is worth discussing what are not the goals Lhotse. Lhotse is not:
\begin{itemize}
    \item A feature extraction library: Lhotse leverages external feature extractors, such as librosa~\cite{mcfee2015librosa}, torchaudio~\cite{torchaudio2021}, kaldifeat~\cite{kaldifeat2021}, etc. Lhotse has utilities for storing and reading pre-computed features from disk or cloud storage; these extractors can also be used on-the-fly.
    \item A data augmentation library: when possible, Lhotse leverages external implementations such as SoX~\cite{sox2021} bindings in torchaudio for audio effects. It is straightforward to use external data augmentation methods with Lhotse, e.g., by including them in a custom \texttt{Dataset} class or passing them via \texttt{augment\_fn} argument in some feature computation methods.
    \item A framework for training: we avoid pushing the users into structuring their training code in some specific way (such as: YAML configuration, Trainer classes, callbacks, inheritance, etc.). We believe it is better to provide a set of tools (manifests, CutSet, transforms, samplers) and examples that make it easier to build data pipelines of arbitrary complexity. 
\end{itemize}

\section{Speech data handling for deep learning}
\label{sec:deep_learning}

\textbf{Standard PyTorch data API.} Lhotse integrates with PyTorch by adopting the PyTorch data API abstractions: \texttt{Dataset}, \texttt{Sampler}, and \texttt{DataLoader}. In a typical PyTorch setup, \texttt{Dataset} "owns" all the data and when queried about specific example indexes, returns tensors describing the model inputs and labels. \texttt{Sampler} is usually created implicitly and for each batch provides a number of example indexes to prepare. \texttt{DataLoader} passes the indexes from \texttt{Sampler} and distributes them to a number of background \texttt{Dataset} workers, and collects and collates the gathered tensors into a batch.

\begin{figure}
    \centering
    \includegraphics[width=\linewidth]{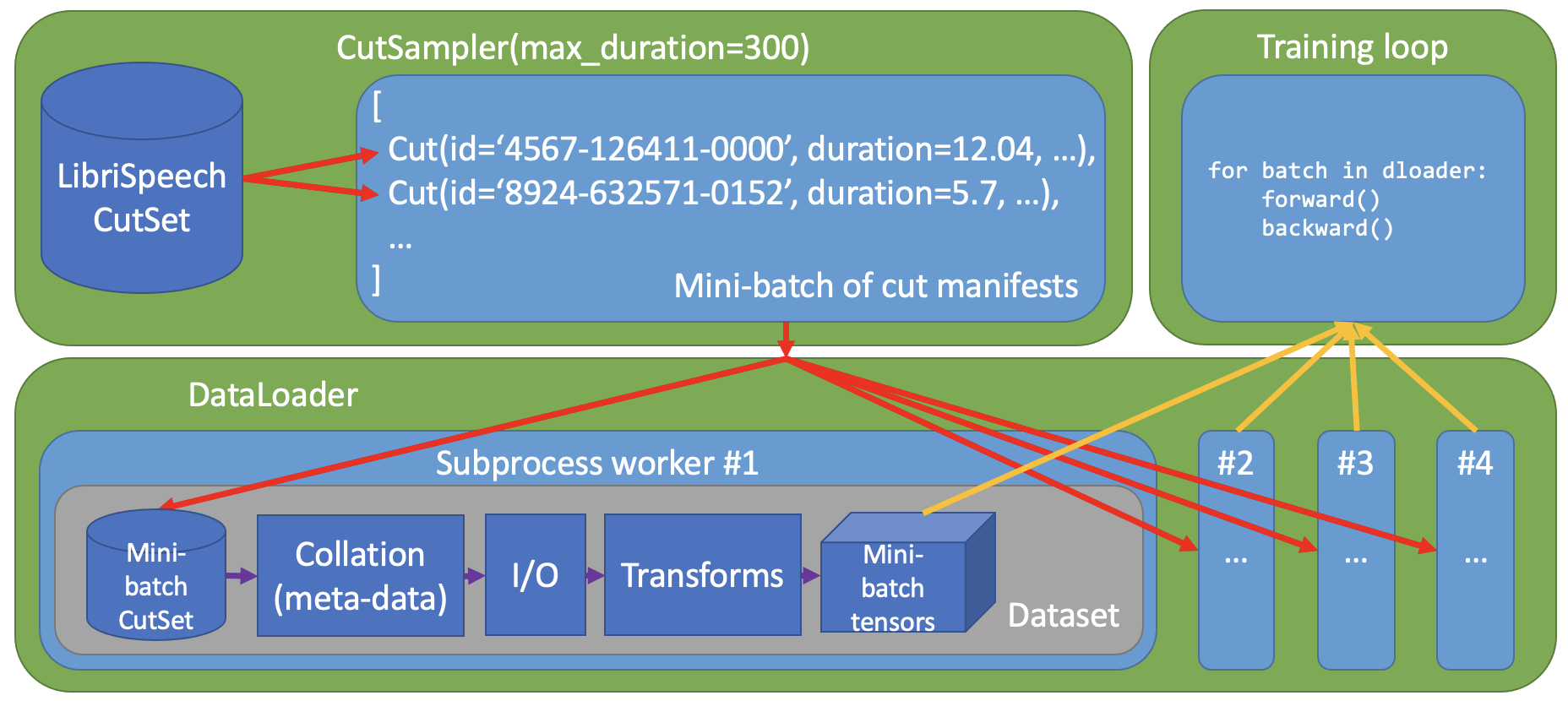}
    \caption{Lhotse implementation of the PyTorch data API.}
    \label{fig:lhotse_pytorch}
\end{figure}

\textbf{Lhotse data API extensions.} Lhotse generally adheres to the aforementioned workflow, illustrated in Figure~\ref{fig:lhotse_pytorch}, with a few notable differences:
\begin{itemize}
    \item \texttt{Dataset} doesn't "own" the data manifests--it simply acts as a function that transforms a mini-batch \texttt{CutSet} into a mini-batch tensor. It performs collation on the meta-data level, possibly re-ordering and concatenating cuts, and only then performs the I/O and transforms.
    \item \texttt{Sampler} (called \texttt{CutSampler}) "owns" the \texttt{CutSet}s with training data and samples individual manifests from it. The batch size is determined dynamically, based on constraints such as the maximum total speech duration (or number of frames) in a mini-batch, or the number of cuts.
\end{itemize}

\textbf{Sampling utterances for training.} Lhotse implements several \texttt{CutSampler}s that are applicable in a wide array of speech processing tasks. \texttt{SingleCutSampler} is used in typical setups where the model works with single utterances (e.g., speech recognition or speech synthesis). \texttt{CutPairsSampler} uses two \texttt{CutSet}s with matching cut IDs and is applicable when the model targets are also speech utterances (e.g., voice conversion or speech translation). \texttt{BucketingSampler} augments previously mentioned sampler types by stratifying the data into similar-duration buckets to minimize the padding, increasing training efficiency. \texttt{ZipSampler} draws batches from multiple samplers and combines them together, which is useful in diversifying the data from multiple sources (domains) during the training. Very large datasets (tens of thousands of hours) are seamlessly handled with minimal memory usage by reading the \texttt{CutSet} manifest lazily and approximately shuffling it on-the-fly if needed (the lazy mode's trade-off is in losing random element access). Lhotse's samplers handle distributed training in the same way as PyTorch's \texttt{DistributedSampler}.

\textbf{Arbitrary task construction.} To support an arbitrary conceivable training setup, the greatest flexibility is required inside the \texttt{Dataset} class that transforms the raw data and meta-data into tensors. Because of that, Lhotse only provides examples of \texttt{Dataset} implementations for speech tasks such as speech recognition, synthesis, voice activity detection, or diarization. The users are encouraged to take these as starting points and adapt them to their use-cases, which may take as little as a few lines of code. We believe that creating a "universal" \texttt{Dataset} class with many points of customization is cumbersome and difficult to understand for most readers, even if possible to achieve. Notably, Lhotse's \texttt{Dataset}s and \texttt{Sampler}s work out-of-the-box with PyTorch's standard \texttt{DataLoader}.

\section{Conclusions}
\label{sec:conclusions}

We presented Lhotse, one of three libraries in the next-generation Kaldi speech processing framework. We showed how Lhotse addresses the major pain points related to speech data handling: it provides a common data representation format for speech corpora through a clean Pythonic API. The library offers off-the-shelf implementations of techniques commonly used in speech processing--such as data augmentation, feature extraction, padding, bucketing, variable-length sequence collation, and others. Practitioners will find more than 30 data preparation recipes for frequently used speech corpora that are stored in a common manifest format. We believe that Lhotse significantly lowers the barrier of entry into speech modelling for machine learning practitioners from other fields. 

Lhotse is open-source, licensed with convenient Apache-2.0 license, and open for community contributions. The library is available at \url{https://github.com/lhotse-speech/lhotse} and PyPI (\texttt{pip install lhotse}).

\begin{ack}
This work is supported by the NSF CCRI grant award number 2120435.
\end{ack}


\bibliography{main}
\bibliographystyle{plain}






\section*{Checklist}

The checklist follows the references.  Please
read the checklist guidelines carefully for information on how to answer these
questions.  For each question, change the default \answerTODO{} to \answerYes{},
\answerNo{}, or \answerNA{}.  You are strongly encouraged to include a {\bf
justification to your answer}, either by referencing the appropriate section of
your paper or providing a brief inline description.  For example:
\begin{itemize}
  \item Did you include the license to the code and datasets? \answerYes{See Section~\ref{sec:conclusions}.}
  \item Did you include the license to the code and datasets? \answerNo{The code and the data are proprietary.}
  \item Did you include the license to the code and datasets? \answerNA{}
\end{itemize}
Please do not modify the questions and only use the provided macros for your
answers.  Note that the Checklist section does not count towards the page
limit.  In your paper, please delete this instructions block and only keep the
Checklist section heading above along with the questions/answers below.

\begin{enumerate}

\item For all authors...
\begin{enumerate}
  \item Do the main claims made in the abstract and introduction accurately reflect the paper's contributions and scope?
    \answerYes{}
  \item Did you describe the limitations of your work?
    \answerYes{} (See the last paragraph of Section~\ref{sec:main_concepts})
  \item Did you discuss any potential negative societal impacts of your work?
    \answerNA{} (we do not expect any)
  \item Have you read the ethics review guidelines and ensured that your paper conforms to them?
    \answerYes{}
\end{enumerate}

\item If you are including theoretical results...
\begin{enumerate}
  \item Did you state the full set of assumptions of all theoretical results?
    \answerNA{}{}
	\item Did you include complete proofs of all theoretical results?
    \answerNA{}
\end{enumerate}

\item If you ran experiments...
\begin{enumerate}
  \item Did you include the code, data, and instructions needed to reproduce the main experimental results (either in the supplemental material or as a URL)?
    \answerNA{}
  \item Did you specify all the training details (e.g., data splits, hyperparameters, how they were chosen)?
    \answerNA{}
	\item Did you report error bars (e.g., with respect to the random seed after running experiments multiple times)?
    \answerNA{}
	\item Did you include the total amount of compute and the type of resources used (e.g., type of GPUs, internal cluster, or cloud provider)?
    \answerNA{}
\end{enumerate}

\item If you are using existing assets (e.g., code, data, models) or curating/releasing new assets...
\begin{enumerate}
  \item If your work uses existing assets, did you cite the creators?
    \answerYes{}
  \item Did you mention the license of the assets?
    \answerYes{}
  \item Did you include any new assets either in the supplemental material or as a URL?
    \answerYes{}
  \item Did you discuss whether and how consent was obtained from people whose data you're using/curating?
    \answerNA{} (Lhotse, as a data tool, only provides utilities for working with existing datasets. The licensing issues are handled by the original dataset creators.)
  \item Did you discuss whether the data you are using/curating contains personally identifiable information or offensive content?
    \answerNA{} (see above)
\end{enumerate}

\item If you used crowdsourcing or conducted research with human subjects...
\begin{enumerate}
  \item Did you include the full text of instructions given to participants and screenshots, if applicable?
    \answerNA{}
  \item Did you describe any potential participant risks, with links to Institutional Review Board (IRB) approvals, if applicable?
    \answerNA{}
  \item Did you include the estimated hourly wage paid to participants and the total amount spent on participant compensation?
    \answerNA{}
\end{enumerate}

\end{enumerate}





\end{document}